\documentstyle{neu}
\begin{document}

\title{NEUTRON STAR/SUPERNOVA REMNANT ASSOCIATIONS}

\author{Victoria M. KASPI \\
{\it    Department of Physics and Center for Space Research,
Massachusetts Institute of Technology, Cambridge, MA, 02139, USA,
vicky@space.mit.edu}
}
\maketitle

\section*{Abstract}

We summarize the current observational
evidence for associations between supernova remnants and neutron
stars, including radio pulsars, proposed ``inactive'' young neutron
stars, ``anomalous'' X-ray pulsars, and soft gamma-ray repeaters, and
argue that the paradigm that all young neutron stars are Crab-like
requires reconsideration.

\section{Introduction}

When Baade \& Zwicky proposed in 1934 that collapsed stars composed of
neutrons could be formed in supernova explosions, they had little
notion of how such creatures would manifest themselves
observationally.  The surprise discovery of pulsars, and in
particular, of the Crab and Vela pulsars in their respective supernova
remnants, heralded the first visual image of the isolated neutron
star: that of a compact, highly magnetized (surface field $\sim
10^{12}$~G) star, spinning down slowly due to magnetic braking,
emitting a collimated beacon, and exciting its surroundings via the
injection of ultra-relativistic particles.  However, even the simplest
follow-up questions, such as whether all neutron stars form in
supernovae, what fraction of supernovae produce neutron stars, and in
particular, whether all young pulsars are born with properties like
those of the Crab and Vela pulsars remain to this day naggingly
unanswered.

Here I summarize observations of neutron stars plausibly
associated with supernova remnants.  The last similar review was
published by Helfand \& Becker (1984).  Their interesting
synthesis of the observational data is beyond the scope of this paper.
More recent reviews including only radio pulsar/supernova remnant
associations can be found elsewhere (Kaspi 1996; Kaspi 1998).

\section{Rotation-Powered Neutron Star/Supernova Remnant
Associations}

Associations between remnants and neutron stars have traditionally
been identified with detections of radio pulsars.  The Helfand \&
Becker (1984) review included only three such objects: the Crab, Vela,
and PSR B1509$-$58.  Since then, the number of associations has
blossomed to between 7 and 23, depending on the one's criteria for
certainty.  The great increase is
for several reasons.  Many young pulsars were found in radio
pulsar surveys of the Galactic plane at observing frequencies near
1400~MHz (Clifton \& Lyne 1986; Johnston et al. 1992), where
dispersion-measure and scatter-broadening of radio pulses are reduced 
relative to traditional pulsar survey frequencies,
like 430~MHz.  Also some, but not all, radio pulsar searches
specifically targeting remnants have been successful (Manchester,
D'Amico \& Tuohy 1985; Kaspi et al. 1996; Gorham
et al. 1996; Lorimer, Lyne \& Camilo 1998).  Finally major advances in
the capabilities of X-ray observatories have led to several recent
exciting discoveries, discussed further below (\S\ref{sec:psr})

Proposed associations may be merely a result of coincidental
projection of the pulsar and remnant on the sky; the probability
for chance alignment is generally significant, particularly for the inner Galactic
plane.  The burden of proof therefore rests on the
observer.  Evidence comes from consideration of whether independent
distance and age estimates agree, whether there is an interaction
between the pulsar and remnant, whether the transverse velocity implied by
the angular displacement of the pulsar from the approximate remnant
geometrical center, its (perhaps naively) assumed birthplace, is
consistent with the known pulsar velocity distribution, and, even
better, if the measured pulsar velocity's magnitude and
direction is as predicted.
For most proposed associations, few of these questions
have clear answers (see Kaspi 1998 for more detail).

Table~1 summarizes proposed rotation-powered pulsar/supernova remnant
associations.  In the Table, PSR and SNR are the pulsar and remnant
names, T is the remnant type (S is shell, P is plerion, C is
composite), $\tau$ and $d$ are the best age and distance estimates for
PSR/SNR, $\beta$ is the angular displacement of the pulsar from the
apparent remnant center in remnant radii, $v_t$ is the best-guess
transverse velocity assuming the pulsar characteristic age and the
most reliable distance estimate, with $v_t$ in bold actual
measurements.  ${\cal E}$ is a rough figure of merit, where ${\cal E}
= 1$ is secure, and ${\cal E} = 5$ dubious.  Note that often ${\cal
E}$ is high for lack of supporting, rather than conclusively damaging,
evidence.

Conclusions to be drawn from Table~1 were discussed by Kaspi (1998).
The main points are that observable young pulsars have larger magnetic
fields, shorter spin periods, but not always larger radio
luminosities, than the typical pulsar.  The evidence from associations
for high pulsar velocities should as yet be considered inconclusive;
proper motion measurements for young pulsars are crucial for deciding
whether many of the associations listed in Table~1 are genuine.  The
detection of bow-shock nebulae via high-resolution radio and X-ray
observations may provide another way to test associations.

\subsection{Recent X-ray Discoveries}
\label{sec:psr}

\noindent
{\bf N157B:}
Marshall et al. (1998) have just discovered a 16~ms X-ray pulsar in
the Crab-like supernova remnant N157B in the Large Magellanic Cloud
using the {\it Rossi X-ray Timing Explorer}.  The surprisingly short
period makes it the most rapidly rotating neutron star known that has
not been spun up.  The measured pulsar characteristic age is
$\sim$5~kyr old, making it the fourth youngest pulsar known, and
suggesting that neutron stars can be born spinning significantly
faster than previously thought, possibly as fast as a few
milliseconds.

\medskip
\noindent
{\bf RCW~103:}
Torii et al. (1998) have discovered 69~ms X-ray pulsations from a
source $7^{\prime}$ north of the center of the well-studied, young
shell supernova remnant RCW~103 using {\it ASCA}, finally confirming
an earlier claim by Aoki, Dotani \& Mitsuda (1992) who used {\it GINGA}.
The spin-down rate implied by the difference in periods indicates that
the pulsar is very young, having age $\sim$8~kyr.  The existence of so
young a pulsar near RCW~103 suggests the two might be related, casting
doubt on the famous central point source being the stellar remnant
(see \S\ref{sec:inactive}) An association requires a transverse
velocity of $\sim$800~km~s$^{-1}$ for a distance of 3.3~kpc, deduced
from HI absorption.  The radio-pulsar counterpart has recently been
discovered at the Parkes observatory (Kaspi et al. in preparation).
The dispersion measure suggests a distance to the pulsar of
$\sim$4.5--7~kpc.  Thus the association can be considered only tentative,
and the nature of the central object remains uncertain.

\medskip
\noindent
{\bf PSR J1105$-$6107:} 
The 63~ms radio pulsar PSR~J1105$-$6107 lies almost three remnant
radii from the approximate center of the supernova remnant
G290.1$-$0.8, also called MSH 11$-$61A (Kaspi et al. 1997).  For an
association, under standard assumptions, the pulsar transverse
velocity must be $\sim$650~km~s$^{-1}$.  The recent detection of the
pulsar at X-ray energies (Gotthelf \& Kaspi 1998) is best explained as
arising from a pulsar wind nebula confined by ram-pressure.
High-resolution X-ray imaging can test the association if a bow-shock
morphology is found.

\medskip
\noindent
{\bf G11.2$-$0.3:}
The supernova remnant G11.2$-$0.3 is the possible counterpart of the
event recorded by the Chinese in AD 386 (Strom 1994).  The recent detection of evidence for 65~ms X-ray
pulsations (Torii et al. 1997) is exciting, as after the Crab, this is
the only pulsar associated with an historic event.  From the single
{\it ASCA} observation, $\dot{P} < 8 \times 10^{-13}$.  If the pulsar
was born in AD 386, and assuming a short birth spin period, the
implied $\dot{P} = 6.4 \times 10^{-13}$, consistent with the upper
limit.  This implies a spin-down luminosity $\dot{E} < 9 \times
10^{37}$~erg~s$^{-1}$, and surface magnetic field $B < 1 \times
10^{13}$~G, reasonable for a
young pulsar.  Confirmation  measurement of
$\dot{P}$ are top priorities for the future.

\section{``Inactive'' Neutron Star/Supernova Remnant Associations}
\label{sec:inactive}

Several unidentified X-ray point sources observed in
supernova remnants have been hypothesized as being
neutron stars emitting X-rays from their initial cooling after
formation.  To date, the sources that fall in
this category are: 1E~1207.4$-$5209 in G296.5+10.0 (Mereghetti,
Bignami \& Caraveo 1996), 1E~1613$-$5055 in RCW~103 (Tuohy \& Garmire
1980; but see also \S\ref{sec:psr} and
Torii et al. 1998), RX~J0002+6246 in G117.7+0.6 (Hailey \& Craig
1995), and a point source in Puppis~A (Petre, Becker \& Winkler 1996).
The observational properties of these sources that support the neutron
star identification are: they are X-ray point sources to within
available spatial resolutions, they are coincident with supernova
remnants, they have high X-ray to optical luminosity ratios,
their spectra are roughly consistent with blackbody emission with
temperatures and emitting surface areas roughly as expected for
neutron stars, and their X-ray emission does not vary
long-term.  However they are clearly different from pulsars like the
Crab because no X-ray pulsations are seen, no associated
radio or $\gamma$-ray emission, pulsed or unpulsed, has been detected,
and no evidence for extended pulsar-powered synchrotron nebulae is
observed.  These objects challenge the conventional paradigm that
young neutron stars are born as pulsars, energetic and spinning fast,
exciting synchrotron nebulae in their immediate surroundings.  An
unambiguous conclusion regarding the neutron-star nature of these
objects could come from the detection of low-pulsed-fraction X-ray
pulsations, expected for magnetized, thermally cooling sources, or
through high-resolution spectral X-ray observations, which could
reveal absorption features predicted in models of neutron star
atmospheres (see Pavlov, this volume).

\section{``Anomalous'' X-ray Pulsar/Supernova Remnant Associations}
\label{sec:magnetar}

The class of objects collectively known as ``anomalous'' X-ray pulsars
is characterized by pulsations long in duration compared with those of
the radio pulsar population (5-9~s), and slow but steady spin-down.
The members of this class are 4U~0142+61, RX~J0720.4$-$3125,
1E~1048.1$-$5937, 4U~1626$-$67, RX~J1838.4$-$0301 and 1E~2259+586.
Mereghetti \& Stella (1995) summarize the properties of these systems
and argue that they are binaries in which the neutron star accretes
from so low-mass a companion that the orbital Doppler shift is not
detectable.  No optical counterpart has been found for any of the
sources except 4U~1626$-$67, which is certainly an X-ray binary (Middleditch et al. 1981).

It is intriguing that two of the sources, 1E~2259+586 (Fahlman \&
Gregory 1981) and RX~J1838.4$-$0301 (Schwentker 1994) are associated
with supernova remnants. These associations
are unexplained in the binary model, although they would not represent
the first known X-ray binaries associated with remnants.  That honor
goes to SS~433, a well-known unusual radio-emitting X-ray binary
associated with the supernova remnant W50 (Margon 1984).

The anomalous X-ray pulsars in supernova remnants could be young, isolated, and
rotation-powered.  However their X-ray luminosities far exceed the
implied spin-down luminosities.  This demands the introduction of a
physical mechanism for the production of X-rays not previously
observed in neutron stars.  Furthermore, the implied surface magnetic
fields are enormous, $\sim 10^{14}$~G.  This has led to them being
dubbed ``magnetars,'' and suggests that the origin of the observed
X-rays could lie, for example, in the decay of the magnetic field by
diffusive processes (e.g. Thompson \& Duncan 1996).

\medskip
\noindent
{\bf Kes 73:}
The recent discovery of a 12~s X-ray pulsar in the young supernova
remnant Kes~73 (Vasisht \& Gotthelf 1997) adds renewed strength to the
magnetar hypothesis.  The strategic location of the pulsar near the
center of the remnant makes a chance coincidence improbable.  A weak
archival {\it ROSAT} detection of the periodicity implies a spin-down
rate and hence $\dot{E}$ and $B$ that are consistent with the magnetar
model.  Confirmation of the spin-down rate and long-term timing are
important for establishing the nature of this source.

\section{Soft Gamma-Ray Repeaters}

The class of objects known as ``soft gamma-ray repeaters'' includes
three sources: SGRs 0526$-$66, 1806$-$20, and 1900+14.  They
are characterized by recurrent episodes of soft-spectrum,
short-duration $\gamma$-ray emission (see Kouveliotou 1996 for a review).
Multi-wavelength studies suggest that they represent
yet another facet of the young neutron star population, since two of
the three (SGRs 0526$-$66 and 1806$-$20) are apparently associated with
supernova remnants (Cline et al 1982; Kulkarni \& Frail 1993).  The
absence of an obvious remnant host to SGR~1900+14 remains problematic
(Vasisht et al. 1996). Thompson \& Duncan (1996) discuss SGRs in the
context of a magnetar model, suggesting that the SGR
phenomenon may represent a phase of evolution in the life of a
magnetar, with X-ray pulsations (\S\ref{sec:magnetar}) characterizing
a different stage.  
SGRs are discussed by Kulkarni in more detail elsewhere in this volume.

\section{Conclusions}

The most striking conclusion following examination of the associations
between neutron stars and supernova remnants is that the paradigm that
every young neutron star is like the Crab pulsar requires
reconsideration.  Neutron stars, it seems, come in many flavors,
perhaps some yet to be discovered.  The continued analysis of valuable
archival high-energy data, the upcoming launch of the {\it Advanced
X-ray Astrophysics Facility},
as well as the ongoing Parkes multi-beam and Nan\c{c}ay
Galactic plane survey should ensure significant progress, and even
perhaps more surprises in this field in the near future.

\section{References}


\re
 Anderson,~S. et al., 1996, ApJ, 468, L55
\re
 Aoki, T., Dotani, T., Mitsuda, K. 1992, IAUC 5588
\re
 Caraveo,~P.~A., 1993, ApJ, 415, L111
\re
 Caswell,J., Kesteven,M., Stewart,R., Milne,D., Haynes,R., 1992,
 ApJ, 399, L151
\re
 Clifton,~T.~R., Lyne,~A.~G., 1986, Nature, 320, 43
\re
 Cline, T., et al., 1982, ApJ, 255, L45
\re
 Damashek,~M., Taylor,~J.~H., Hulse,~R.~A., 1978, ApJ, 225, {L}31
\re
 Davies,~J.~G., Lyne,~A.~G., Seiradakis~J.~H., 1972, Nature, 240, 229
\re
 Fahlman,~G.~G., Gregory,~P.~C., 1981, Nature, 293, 202
\re
 Finley,~J.~P., \"{O}gelman,~H., 1994, ApJ, 434, L25
\re
 Finley, J. P., Srinivasan, R. \& Park, S., 1996, ApJ, 466, 938
\re
 Frail,~D.~A., Kulkarni,~S.~R., 1991, Nature, 352, 785
\re
 Frail,~D.~A., Kulkarni,~S.~R., Vasisht,~G., 1993, Nature, 365, 136
\re
 Frail,~D.~A., Goss,~W.~M., Whiteoak,~J. B.~Z., 1994, ApJ, 437, 781
\re
 Gaensler,~B.~M., Johnston,~S., 1995, MNRAS, 275, 73P
\re
 Gorham,~P., Ray,~P., Anderson~S., Kulkarni,~S., Prince,~T., 1996,
 ApJ, 458, 257
\re
 Gotthelf, E. V., Kaspi, V. M., 1998, ApJL, in press
\re
 Hailey, C. J., Craig, W. W., 1995, ApJ, 455, L151
\re
 Helfand, D. J., Becker, R. H. 1984, Nature, 307, 215 
\re
 Johnston,~S., et al., 1992, MNRAS, 255, 401
\re
 Johnston,S.,Manchester,R.,Lyne,A.,Kaspi,V.,D'Amico,N., 1995, A\&A, 293, 795
\re
 Kaspi,V.,Manchester,R.,Johnston,S.,Lyne,A.,D'Amico,N., 1992, ApJ, 399, L155
\re
 Kaspi,~V.~M. et al. 1993, ApJ, 409, L57
\re
 Kaspi, V. M., 1996, {\it IAU Colloquium 160: Pulsars:
 Problems and Progress}, ASP Conference Series, 105, 375
\re
 Kaspi,V., Manchester,R., Johnston,S., Lyne,A., D'Amico,N., 1996,
 AJ, 111, 2028
\re
 Kaspi,~V.~M. et al. 1997, ApJ, 485, 820
\re
 Kaspi, V. M., 1998, Adv. Space. Res., 21, in press
\re
 Kassim,~N.~E., Weiler,~K.~W., 1990, Nature, 343, 146
\re
 Kouveliotou, C., 1996, {\it IAU Symp. 165: Compact Stars in
 Binaries},  Kluwer, Dordrecht, 477
\re
 Kulkarni, S. R. et al., 1988, Nature, 331, 50
\re
 Kulkarni,~S., Predehl,~P., Hasinger,~G., Aschenbach,~B., 1993, Nature, 362, 135
\re
 Kulkarni,~S.~R., Frail,~D.~A., 1993, Nature, 365, 33
\re
 Large,~M.~I., Vaughan,~A.~E. \& Mills, B. Y., 1968, Nature, 220, 340 
\re
 Large,~M.~I., Vaughan,~A.~E., 1972, Nature Phys. Sci., 236, 117
\re
 Leahy,~D.~A., Roger,~R.~S., 1991, Astron. J., 101, 1033
\re
 Lorimer, D. R., Lyne, A. G., Camilo, F. 1998, A\&A, 331, 1002
\re
 Lyne,~A.~G., Lorimer,~D.~R., 1994, Nature, 369, 127
\re
 Manchester,~R. N., D'Amico,~N., Tuohy,~I.~R., 1985, MNRAS, 212, 975
\re
 Manchester,R.,Kaspi,V.,Johnston,S.,Lyne,A.,D'Amico,N., 1991, MNRAS, 253, 7P
\re
 Margon, B. 1984, ARAA, 22, 507
\re
 Marshall, F., Middleditch, J. Zhang, W., Gotthelf, E., 1998, IAUC 6810
\re
 McAdam,~W.~B., Osborne,~J.~L., Parkinson,~M.~L., 1993, Nature, 361,
 516
\re
 Mereghetti, S., Stella, L. 1995, 442, L17
\re
 Mereghetti, S., Bignami, G. F., Caraveo, P. A., 1996, ApJ, 464, 842
\re
 Middleditch, J., Mason, K. O., Nelson, J. E., White, N. E. 1981, ApJ,
 244, 1001
\re
 Nicastro, L., Johnston, S. \& Koribalski, B. 1996, A\&A, 306, 49
\re
 Petre~R., Becker~C.~M., Winkler~P.~F., 1996, ApJ, 465, L43
\re
 Phillips,~J.~A., Onello,~J.~S., 1993, {\it Massive Stars: Their Lives in the Interstellar
 Medium}, ASP, 35, ~419
\re
 Ray,~P.~S. et al., 1996, ApJ, 470, 1103
\re
 Routledge,~D., Vaneldik,~J.~F., 1988, ApJ, 326, 751
\re
 Schwentker, O. 1994, A\&A, 286, L47
\re
 Seward, F. D., Harnden, F. R., 1982, ApJ, 256, L45
\re
 Seward,~F.~D., Harnden,~F.~R., Helfand~D.~J., 1984, ApJ, 287, L19
\re
 Staelin,~D.~H., Reifenstein,~{III}~E.~C., 1968, Science, 162, 1481
\re
 Strom, R. G. 1994, A\&A, 288, L1
\re
 Thompson,~R.~J., C\'ordova~F.~A., 1994, ApJ, 421, L13
\re
 Thompson, C., Duncan, R. C. 1996, ApJ, 473, 322
\re
 Torii, K., Tsunemi, H., Dotani, T., Mitsuda, K. 1997, ApJ, 489, L145
\re
 Torii, K. et al 1998, ApJ, 494, L207
\re
 Tuohy, I., Garmire, G., 1980, ApJ, 239, 107
\re
 Vasisht, G., Kulkarni, S. R., Frail, D. A., Greiner, J. 1996, ApJ,
 456, L59
\re
 Vasisht, G., Gotthelf, E. V. 1997, ApJ, 486, L129
\re
 Wolszczan,~A., Cordes,~J.~M., Dewey,~R.~J., 1991, ApJ, 372, L99


\footnotesize
\begin{table}[t]
    \caption{Proposed Rotation-Powered
    Pulsar/Supernova Remnant Associations.} 
\begin{center}
\begin{tabular}{lccccrrrc}\hline
\multicolumn{1}{c}{PSR} & SNR & Type & \multicolumn{1}{c}{$\tau$} & \multicolumn{1}{c}{$d$} & $\beta$ &
\multicolumn{1}{c}{$v_t$} & $\cal E$ & Refs \\
    &     &      & \multicolumn{1}{c}{(kyr)} & \multicolumn{1}{c}{(kpc)} &         &\multicolumn{1}{c}{(
km/s)} &    \\ \hline
B0531+21 & Crab & P & 1.3/0.9 & 2/2 & $\sim$0 & {\bf 125} & 1 & 1 \\
B0540$-$69 & SNR0540$-$693 & P & 1.7/0.6 & 50/50 & $\sim$0 & $\sim$0 &
1 & 2 \\
J0537$-$6910 & N157B & C & 5/4 & 50/50 & $\sim$0.5 & $\sim$600 & 1 &38\\
B0833$-$45 & Vela & C & 11/18 & 0.6/0.5 & 0.3 & {\bf 170} & 1 & 32 \\
B1509$-$58 & MSH 15--52 & C & 1.7/10 & 5.7/4.2 & 0.2 & 3000? & 2 & 27 \\
B1757$-$24 & G5.4$-$1.2 & C & 16/14 & 4.6/5 & 1.2 & 1600 & 2 & 5,6\\
B1853+01 & W44 & S & 20/$\sim$10 & 3/3.1 & 0.6 & 250 & 2 & 7 \\
B1951+32 & CTB 80 & C & 107/96 & 2.4/3  & $\sim$0 & 300 & 2 & 8 \\
J1341$-$6220 & G308.8$-$0.1 & C & 12/32 & 8.7/7 & 0.35 & 600 & 3 & 3,4 \\
B1800$-$21 & G8.7$-$0.1 & S & 16/15-28 & 4/3.2-4.3 & $\sim$0 & $\sim$0 & 3 & 33,34 \\
B1823$-$13 & X-ray nebula & P & 21/? & 2.5/? & $\sim$0 & $\sim$0 & 3 &
28 \\
J0538+2817 & S147 & S & 600/80-200 & 1.8/0.8--1.6 & 0.4 & 30 & 3 & 26 \\
B1643$-$43 & G341.2+0.9 & C & 33/- & 6.9/8.3-9.7 & 0.7 & 475 & 3 & 29 \\
B2334+61 & G114.3+0.3 & C & 41/10-100 & 2.4/1.8 & 0.1 & $<$50 & 3 & 9 \\
B1758$-$23 & W28 & C & 58/35-150 & 13.5/2 & 1.0 & 200 & 3 & 10,11\\
J1811.5$-$1926 & G11.2$-$0.3 & C & -/2 & -/$>$5 & $<$1 & ? & 4 & 35 \\
B1610$-$50 & Kes 32 & S & 7.5/5 & 7/3-7 & 1.5 & 1600 & 4 & 12,13 \\
J1617$-$5055 & RCW~103 & S & 8/4 & 7/3.3 & 1.5 & 800 & 4 & 36,37\\
B1727$-$33 & G354.1+0.1 & ? & 26/- & 4.2/- & $\sim$0 & 460 & 4 & 29 \\
J1105$-$6107 & G290.1$-$0.8 & S & 63/? & 7/7 & 2.9 & 650 & 4 & 25 \\
B1830$-$08 & W41 & S & 148/$<$50 & 4-5/4.8 & 1.6 & 200 & 4 & 14,15 \\
B1855+02 & G35.6$-$0.5 & ? & 160/- &  9/4 or 12 & 0.4 & 100 & 4 & 16 \\
J1627$-$4845 & G335.2+0.1 & S & 2700/- & 6.8/6.5 & 0.4 & 70 & 4 & 17 \\
B1706$-$44 & G343.1$-$2.3 & C & 17.5/- & 2.4-3.2/3 & 1.0 & {\bf $<$50} & 4 & 29,30,31 \\
B1930+22 & G57.3+1.2 & ? & 40/- & 9.6/4.5  & 0.5 & 750 & 5 & 18 \\
B0611+22 & IC 443 & S & 89/65 & 4.7/1.5 & 1.7 & {\bf 110} & 5 & 19 \\
B0656+14 & Monogem & S? & 110/60-90 & 0.8/0.3 & 0.5  & {\bf 250} & 5 & 20 \\
B1832$-$06 & G24.7+0.6 & C & 120/12 & 6.3/4.4 & 1.6 & 360 & 5 & 15 \\
J2043+2740 & Cygnus Loop & S & 1200/20 & 1.1/0.6 & 2.5 & 1500 & 5 & 21 \\
B1154$-$62 & G296.8$-$0.3 & S & 1600/25 & 10/4 & 1.4 & 550 & 5 & 22 \\
B0458+46 & G160.9+2.6 & S & 1800/30-100 & 1.8/1-4 & 0.3 & {\bf $<$150} & 5 & 23,24 \\\hline
\end{tabular}
\end{center}
\noindent
Refs: [1] Staelin \& Reifenstein (1968) \nocite{sr68}
[2] Seward et al. (1984) \nocite{shh84}
[3] Kaspi et al. (1992), \nocite{kmj+92}
[4] Caswell et al. (1992) \nocite{cks+92}
[5] Frail \& Kulkarni (1991) \nocite{fk91}
[6] Manchester et al. (1991) \nocite{mkj+91}
[7] Wolszczan et al. (1991) \nocite{wcd91}
[8] Kulkarni et al. (1988) \nocite{kcb+88}
[9] Kulkarni et al. (1993) \nocite{kpha93}
[10] Kaspi et al. (1993) \nocite{klm+93}
[11] Frail et al. (1993) \nocite{fkv93}
[12] Caraveo (1993) \nocite{car93}
[13] Johnston et al. (1995) \nocite{jml+95}
[14] Clifton \& Lyne (1986) \nocite{cl86}
[15] Gaensler \& Johnston (1995a) \nocite{gj95a}
[16] Phillips \& Onello (1992) \nocite{po93}
[17] Kaspi et al. (1996) \nocite{kmj+96}
[18] Routledge \& Vaneldik (1988) \nocite{rv88}
[19] Davies, Lyne \& Seiradakis (1972) \nocite{dls72}
[20] Thompson \& Cordova (1994)  \nocite{tc94}
[21] Ray et al. (1996) \nocite{rtj+96}
[22] Large \& Vaughan (1972) \nocite{lv72}
[23] Damashek et al. (1978) \nocite{dth78}
[24] Leahy \& Roger (1991) \nocite{lr91}
[25] Kaspi et al. (1997) \nocite{kbm+97}
[26] Anderson et al. (1996) \nocite{acj+96}
[27] Seward \& Harnden (1982) \nocite{sh82}
[28] Finley, Srinivasan \& Park (1996) \nocite{fsp96}
[29] Frail, Goss, \& Whiteoak (1994)
[30] McAdam, Osborne \& Parkinson (1993)
[31] Nicastro et al. (1996)
[32] Large, Vaughan \& Mills (1968)
[33] Kassim \& Weiler (1990)
[34] Finley \& \"{O}gelman (1994)
[35] Torii et al. (1997)
[36] Torii et al. (1998)
[37] Kaspi et al. (1998)
[38] Marshall et al. (1998)
\end{table}
\normalsize

\end{document}